# Independence of topological surface state and bulk conductances in three-dimensional topological insulators

Shu Cai[1,4], Jing Guo[1], Vladimir A. Sidorov[3], Yazhou Zhou[1], Honghong Wang[1,4], Gongchang Lin[1,4], Xiaodong Li[5], Yanchuan Li[5], Ke Yang[6], Aiguo Li[6], Qi Wu[1], Jiangping Hu[1,4], S. K. Kushwaha[2], Robert J Cava[2]†, and Liling Sun[1,4]†

[1]*Institute of Physics and Beijing National Laboratory for Condensed Matter Physics, Chinese Academy of Sciences, Beijing 100190, China*
[2]*Department of Chemistry, Princeton University, Princeton, New Jersey 08544, USA*
[3]*Institute for High Pressure Physics, Russian Academy of Sciences, 142190 Troitsk, Moscow, Russia*
[4]*University of Chinese Academy of Sciences, Beijing 100190, China*
[5]*Institute of High Energy Physics, Chinese Academy of Sciences, Beijing 100049, China*
[6]*Shanghai Synchrotron Radiation Facilities, Shanghai Institute of Applied Physics, Chinese Academy of Sciences, Shanghai 201204, China*

The archetypical 3D topological insulators $Bi_2Se_3$, $Bi_2Te_3$ and $Sb_2Te_3$ commonly exhibit high bulk conductivities, hindering the characterization of the surface state charge transport. The optimally doped topological insulators $Bi_2Te_2Se$ and $Bi_{2-x}Sb_xTe_2S$, however, allow for such characterizations to be made. Here we report the first experimental comparison of the topological surface states and bulk conductances of $Bi_2Te_2Se$ and $Bi_{1.1}Sb_{0.9}Te_2S$, based on temperature-dependent high-pressure measurements. We find that the surface state conductance at low temperatures remains constant in the face of orders of magnitude increase in the bulk state conductance, revealing in a straightforward way that the topological surface states and bulk states are decoupled at low temperatures, consistent with theoretical models, and confirming topological insulators to be an excellent venue for studying charge transport in 2D Dirac electron systems.

Topological insulators (TIs) are, theoretically, quantum materials that display a bulk band gap like an ordinary insulator, but a conducting surface state that is topologically protected due to a combination of spin-orbit interactions and time-reversal symmetry[1-6]. The topologically non-trivial nature of the spin-helical Dirac fermion surface states in TIs has attracted wide interest in the research community because it results in rich new physics and materials that have potential applications in quantum technology[7,8].

An ideal TI should have topologically-protected metallic surface states (TSS), but also an electrically insulating bulk. Unfortunately, early generations of bulk TIs, materials such as the $M_2X_3$ (M=Bi and Sb, X=S, Se and Te) tetradymites, are not typically insulating[2,9-13], which prevented the early characterization of many of the charge transport properties of the surface states. Recently, however, insulating bulk crystals of the tetradymite TI variants $Bi_2Te_2Se$ (BTS)[14-16] and $Bi_{1.1}Sb_{0.9}Te_2S$ (BSTS)[17,18] have been grown; these two materials are high quality topological insulators with very low bulk carrier concentrations[16,18-20]. The bulk bandgaps of BTS and the BSTS are ~310 meV[14,19,20] and ~350 meV[18], and, in the latter material, the Dirac crossing in the TSS band is well isolated from the bulk states. These compounds provide an ideal platform for studying the evolution of the TSS conductance and its connections with the conductance of the bulk state.

Although it is widely accepted that topological surface states and bulk electronic states should act independently, the degree to which they might or might not intermix has not been addressed in a straightforward experiment. We address that shortcoming here, finding that the surface state and bulk charge transport in two topological insulators are truly independent at low temperatures. We use transport measurements under applied pressure to show this, because pressure is a clean way to tune electronic states in materials without introducing chemical complexity. High pressure measurements addressing other issues have been employed in studies on some TIs[21-26], but here we report the first high pressure investigations of insulating BTS and BSTS single crystals that directly address the interdependence of the surface and bulk state conductances. The surprising result is that above temperatures of ~10 K, the bulk and surface states appear to be intermixing and the bulk conductance changes by orders of magnitude, while, below 10 K, the surface state conductance changes very little with temperature or pressure, an indication that they are truly independent at that temperature and below.

Theorists have pointed out since the earliest days that the surface states on TIs are protected by time reversal and spatial inversion symmetry[1,6]. The analysis of quantum oscillations and other characteristics of the electronic states at low temperatures have been consistent with this expectation, but the current measurements show the conditions under which this can be assumed to be true in high quality materials. Logically speaking, TIs require materials with relatively small bulk electronic band gaps due to the importance of band inversion and the impact of spin orbit coupling on the bulk electronic band structure at certain points in the Brillouin zone. Thus at high temperatures electrical transport measurements on all but the thinnest crystals should be dominated by the bulk states, in which charge carriers are present due to thermal excitation. Conversely, only for topological insulators where the bulk behavior is that of a good semiconductor, *i.e.* those with low bulk carrier concentrations, still relatively rare even at this date, should the metallic surface states dominate at low temperatures where they short circuit the high resistance of the bulk states in a sample crystal. In intermediate temperature regimes, mixing of the bulk and surface state charge carriers is expected. The current measurements allow those temperature regimes to be defined experimentally for high quality crystals of optimally doped insulating topological insulators.

Firstly it is important to establish that no structural phase transitions are present in the pressure range of our experiments that would complicate the interpretation of the data. It is known that BTS maintains its tetradymite structure to 8 GPa[24], but there are no reports of the high pressure structure of BSTS. Thus, we performed high pressure X-ray diffraction measurements to characterize the structure of BSTS; no structural phase transition was found up to beyond the pressures of the current measurements (see SI figure S1).

We performed temperature-dependent resistance measurements on the single crystals of BTS and BSTS at low pressures. As shown in Fig.1a and 1b, the resistance of the two materials increases strongly upon cooling and then saturates at ~88 K for BTS and ~155 K for BSTS. The saturation of the resistance occurs below temperatures where the metallic topological surface state resistance "short circuits" the strongly increasing bulk resistance, even for bulk crystal pieces[10,16,27]. We denote the resistance saturation temperature as $T^*$: (Fig.1a and 1b). $T^*$ for both TIs shifts to lower temperature upon increasing pressure (Fig.1c and Fig.S2); this is because the bulk insulating behavior of the two materials is systematically suppressed by applying pressure, through a decrease in the activation energy for bulk conduction $E_a$. To extract an estimate of the $E_a$s, we fitted our higher temperature data by the Arrhenius equation (SI, Fig. S2), showing how $E_a$ for both TIs decreases with increasing pressure (Figs.1d and S2). Thus the decrease in the crossover temperature $T^*$ with pressure is associated with a pressure-induced reduction in the activation energy for the generation of bulk carriers, as the resistance of the bulk goes down with pressure, one has to cool to lower temperatures to get comparable bulk and surface state conductances. We note that $T^*$ for BSTS is higher than that of BTS (Fig.1c), a difference that results from the relative magnitudes of the transport activation energies. Indeed, the bandgap of bulk BSTS is larger than that of bulk BTS (SI, Fig. S2). In addition, $E_a$ for BSTS decreases less strongly under pressure than is seen for BTS, resulting in the fact that $T^*$ for BSTS shows a weaker dependence on pressure than is seen for BTS (Figs. 1c,1d and S2).

To illustrate the effect of pressure on the conductances of the TSS in BTS and BSTS more quantitatively, we describe the resistance-temperature (R-T) behavior of these materials via the commonly-employed two channel model[17,28]. Because conductivities are additive, the total conductance ($G_{tot}$) is then the sum of a thermally activated bulk conductance ($G_{bul}$) and a metallic surface conductance ($G_{sur}$)[28]:

$$G_{tot}(T) = G_{bul}(T) + G_{sur}(T)$$

In this way we can examine $G_{sur}(T)$ by subtracting $G_{bul}(T)$ from our experimental data,

where the semiconducting $G_{bul}(T)$ is simply extrapolated to temperatures lower than our fit temperatures. As shown in Fig.2a and 2b, the black squares are our experimental data, the red solid lines are the data fitted by $R_{bul}(T) = [R_{0bul} exp[(T/T_0)^{-0.25}]$ because this equation can give a significant overlap in the activated temperature range[10,16](SI, Fig. S3), where $G_{bul}(T) = 1/R_{bul}(T)$ and the green triangles are the data after subtraction. The green triangles therefore represent the temperature-dependent conductance of the surface states only. $R_{bul}$ increases exponentially over a wide range of decreasing temperature. Based on these results, we can define two relevant temperatures as $T^*$ and $T'$. Below $T^*$, the surface state dominates the resistance, while above $T'$, the bulk insulating state dominates. Between $T^*$ and $T'$, both surface and bulk carriers contribute in a major way to the total conductance of the samples. Because the two channel model can describe our experimental data well (Fig.2a and 2b), it allows us to extract the pressure and temperature dependent conductivities of both the surface and bulk states for the two material investigated (Fig. 2c and 2d, the details see Fig.S4), in which the equations of $G_{bul}(T) = 1/[R_{0bul} exp[(T/T_0)^{-0.25}]$ and $G_{sur}(T) = 1/[R_0+AT]$ are employed. It is seen that the $G_{bul}$ for both TIs increases dramatically above 100 K, especially under pressure, while in contrast $G_{sur}$ appears to be insensitive to both temperature and pressure. Specifying this general behavior in further detail, the temperature-pressure regime where the surface states and the bulk states dominate the sample resistances, for samples where the ratio of the surface area to the sample thickness is approximately 4680 μm$^2$:12 μm (78×60:12=390:1), are shown in the SI (Fig S5). Note that conductance $G$ goes as $G=(e^2/h)k_F l$, where $e$ is charge, $h$ is Plank constant, $k_F$ is wave vector of Fermi surface and $l$ is the mean free path, then the $k_F l$ product for the surface state in the two materials should have to be the same. Whether there exists a universal $k_F l$ value for the TSS of topological insulators deserves further investigation.

Figures 3 and 4 show a summary of the data that show the independence of the bulk and surface state conductances at low temperatures and pressures. Figure 3 shows the details of the pressure- and temperature-dependent behavior of the conductances for both the bulk and surface states, along with a schematic of the electronic structure of the materials as a guide for interpreting the data. Fig. 3a shows the pressure dependence of the high temperature, bulk dominated behavior, and the low temperature, surface dominated behavior, for both BTS and BSTS. It is found that the conductances obtained at 270 K show a linear increase upon increasing pressure in the logarithmic scales, reflecting a thermal excitation behavior from the bulk. However, the conductances obtained at 10 K remain almost constant at low pressure. The relative insensitivity of the low temperature conductance to pressure is immediately apparent for both TIs, but this is illustrated more explicitly in Fig. 3b, where the same data is plotted as relative change in conductance. Fig. 3b shows that there is very little change in the total sample conductance in BTS and BSTS at low pressures and temperatures, because this is where the surface state conductance dominates. The data imply that the surface state and bulk conductanes are thus independent in this regime, but this can be shown more directly in a different kind of plot (in contrast, at high temperatures, where the bulk dominates the sample conductance changes dramatically with pressure). The temperature dependence of the surface state conductance, extracted via equation of $R_{bul}(T) = [R_{0bul} exp[(T/T_0)^{-0.25}]$, for both BTS and BSTS is shown in Fig. 3c. It is seen clearly that for the BTS sample, one has to go to low temperatures before there is little temperature dependence, and that pressure has a significant impact on the surface state conductance. This is because even for BTS there are intermixing of the topological surface state and bulk state electrons down to about 5 K, likely related to the smaller bulk gap and higher bulk carrier concentration for this material than for BSTS. In contrast, for optimally doped BSTS the surface state conductance is quite insensitive to both pressure and temperature below 100 K, until the pressure has decreased the bulk conductance activation energy sufficiently, by around 4 GPa, for the resulting bulk carriers to interfere with the surface state charge carriers.

The essential character of the surface state conductance relative to applied pressure and the bulk state conductance are shown in Fig. 4. Fig. 4a shows the pressure dependence of the surface state conductances in BTS and BSTS at 10 K, on linear scales. As seen in the other data analysis presented, the surface state conductance in BTS is more sensitive to temperature and pressure than is BSTS. This is not an intrinsic feature of the surface state but rather reflects the more dramatic changes in the bulk transport gap $Ea$, with pressure in BTS than for BSTS. The fact that the pressure-dependent behavior of the surface state conductance in BTS at around 10 K is not intrinsic to the surface state is illustrated nicely by the convergence of the surface state conductance to a single value at low temperatures and pressures.

The same behavior is more pronounced for BSTS. Fig 4b shows a full summary of the results of this study. In this figure, the surface state conductance at 10 K is plotted vs. the bulk state conductance at 270 K for both materials on a log-linear plot. The bulk state conductance at 270 K, taken by direct measurement, is taken as a model-indepedent reflection of the bulk conductance, although the same is clearly illustrated when using the bulk conductances extrapolated to 10 K (*i.e.* Fig. S6). This shows that in the low temperature low pressure regime, where the bulk conductance is smallest, the surface state conductance is independent of both the bulk state conductance and the material studied.

In summary, our experiments reveal the independence of the bulk and topological surface state charge transport in optimally doped topological insulators at temperatures of 10 K and below. This is one of the fundamental characteristics predicted and assumed to be the case in topological insulators. We have done this by studying the effect of pressure on the topological surface state (TTS) and bulk state conductances in the insulating 3D topological insulators $Bi_2Te_2Se$ and $Bi_{1.1}Sb_{0.9}Te_2S$. These two TIs show the same behavior under pressure: the crossover temperatures $T^*$ of the TTS-dominated conductance shifts to lower temperature with increasing pressure due to the pressure-induced metallization of the normally insulating bulk. The experiments show that the conductance of the topological surface states remains constant under pressure, in spite of the fact that the conductance of the bulk states increases by several orders of magnitude. The weak metallic temperature-dependent behavior of the TSS, independent of the conductance of the insulating bulk as its bulk transport gap is closed, demonstrates a remarkably robust independence of the two conductance channels; very little mixing of the bulk surface state charge transport is present at low temperatures. Thus measurements of surface state quantization, such as quantum oscillations, typically performed at temperatures considerably lower than 10 K, can be considered as good probes of the surface state transport when performed on bulk insulating crystals of materials such as BTS and BSTS. The unaffected conductance of the TTS in BTS and BSTS suggests that these materials may have potential applications in quantum computation, as proposed for intrinsic TIs.

**Materials and methods**

High-quality single crystals of BTS and BSTS were grown by the vertical Bridgman method, as described in Ref.18. Before the experiments, the crystals were freshly cleaved to expose pristine basal plane (001) surfaces.

High pressure resistance measurements for the two TIs were performed in a Toroid type high pressure cell[29]. A mixture of glycerin and water, a liquid pressure transmitting medium, was used for these measurements. The pressure was determined by monitoring the shifts of the superconducting transition temperature of pure lead[30].

High-pressure X-ray diffraction (XRD) measurements were performed in a diamond anvil cell at room temperature on beamline 4W2 at the Beijing Synchrotron Radiation Facility and on beamline 15U at the Shanghai Synchrotron Radiation Facility. Diamonds with low birefringence were selected for these XRD measurements. A monochromatic X-ray beam with a wavelength of 0.6199 Å was employed and silicon oil was taken as a pressure-transmitting medium. The pressure for all measurements in the diamond anvil cell was determined by the ruby fluorescence method[31].

**Acknowledgements**

We thank Profs. Hongming Weng, Qianghua Wang, N. Phuan Ong and Dr. Chongchong Le for helpful discussions. R.J.C. would like to thank Gene Mele for a conversation several years ago about the expected behavior of the bulk band gap in topological insulators under pressure, initially motivating this experiment. The work in China was supported by the National Key Research and Development Program of China (Grant No. 2017YFA0302900, 2016YFA0300300 and 2017YFA0303103), the NSF of China (Grants No. 11427805, No. U1532267, No. 11604376), the Strategic Priority Research Program (B) of the Chinese Academy of Sciences (Grant No. XDB07020300). The work at Princeton was supported by the ARO MURI on Topological Insulators, grant W911NF-12-1-0461.

†Correspondence and requests for materials should be addressed to L. Sun (llsun@iphy.ac.cn) or R.J. Cava (rcava@Princeton.EDU)

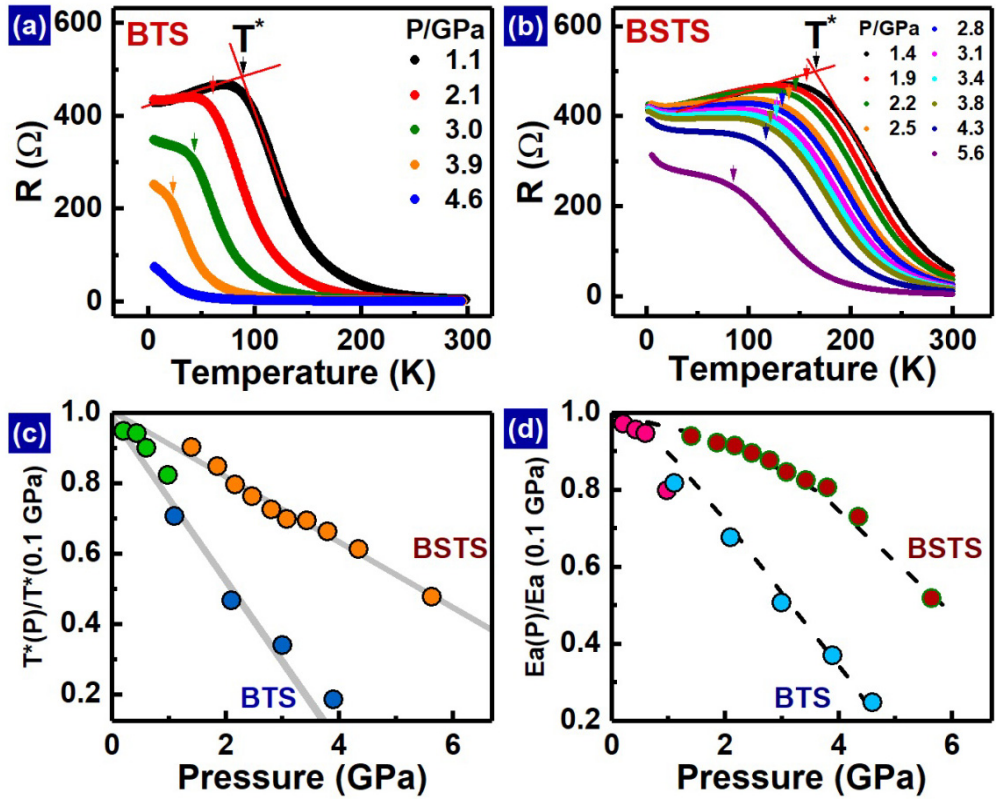

**Figure 1 Transport properties of the 3D topological Insulators BTS and BSTS as functions of temperature and pressure.** (a) and (b) Temperature dependence of the electrical resistance obtained at different pressures for BTS and BSTS, displaying the evolution of *T\**, the temperature that marks the crossover between TSS-dominated and bulk-dominated resistivity. (c) Pressure dependence of the relative behaviors of the crossover temperature, *T\*(P)/T\*(0.1 GPa),* for the two TIs BTS and BSTS, showing a downward trend with increasing pressure. The data labeled by colored circles were obtained from different experimental runs. (d) The relative transport activation energies (*Ea*) as a function of pressure for BTS and BSTS, displaying a strong decrease with pressure. The *T\** and *Ea* values at the lowest pressures are ~125

K and ~144 meV for BTS and ~180 K and ~165 meV for BSTS respectively. Further information can be found in the SI, Fig. S1.

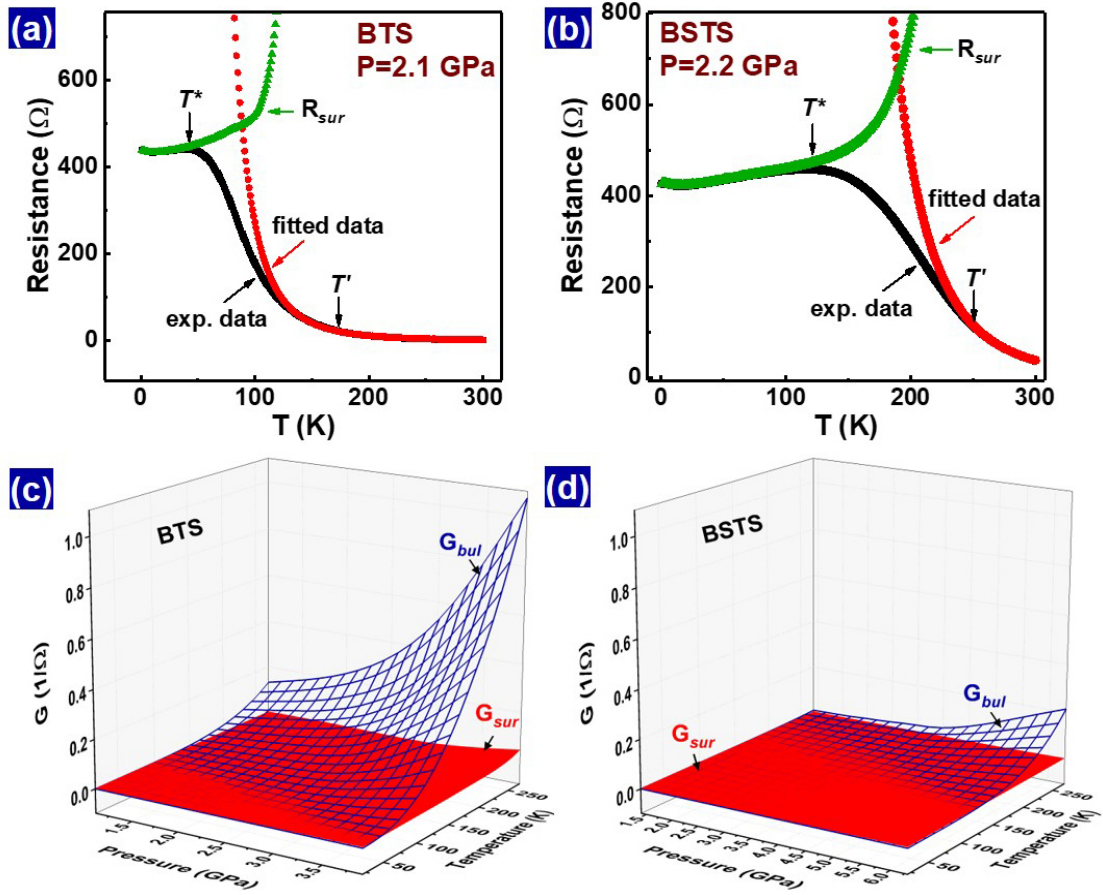

**Figure 2 The temperature and pressure dependent resistivities and conductances for BTS and BSTS.** (a) and (b) Resistance as a function of temperature at fixed pressure for the two materials. The back squares are our experimental data, the red circles are the data fitted for the insulating bulk in the high temperature range by the equation of $R_{bul}(T) = [R_{0bul} exp[(T/T_0)^{-0.25}]$ and the green triangles are the surface state conductance, obtained from our experimental (black) data by subtracting the fitted bulk conductance extrapolated from the high temperature range (red). (c) and (d) The pressure and temperature dependence of the conductance of the TSS and the insulating bulk for BTS and BSTS, showing that the conductance of the bulk ($G_{bul}$) increases with both increasing pressure and temperature, while the conductance of TSS ($G_{sur}$) stays relatively constant with pressure and temperature.

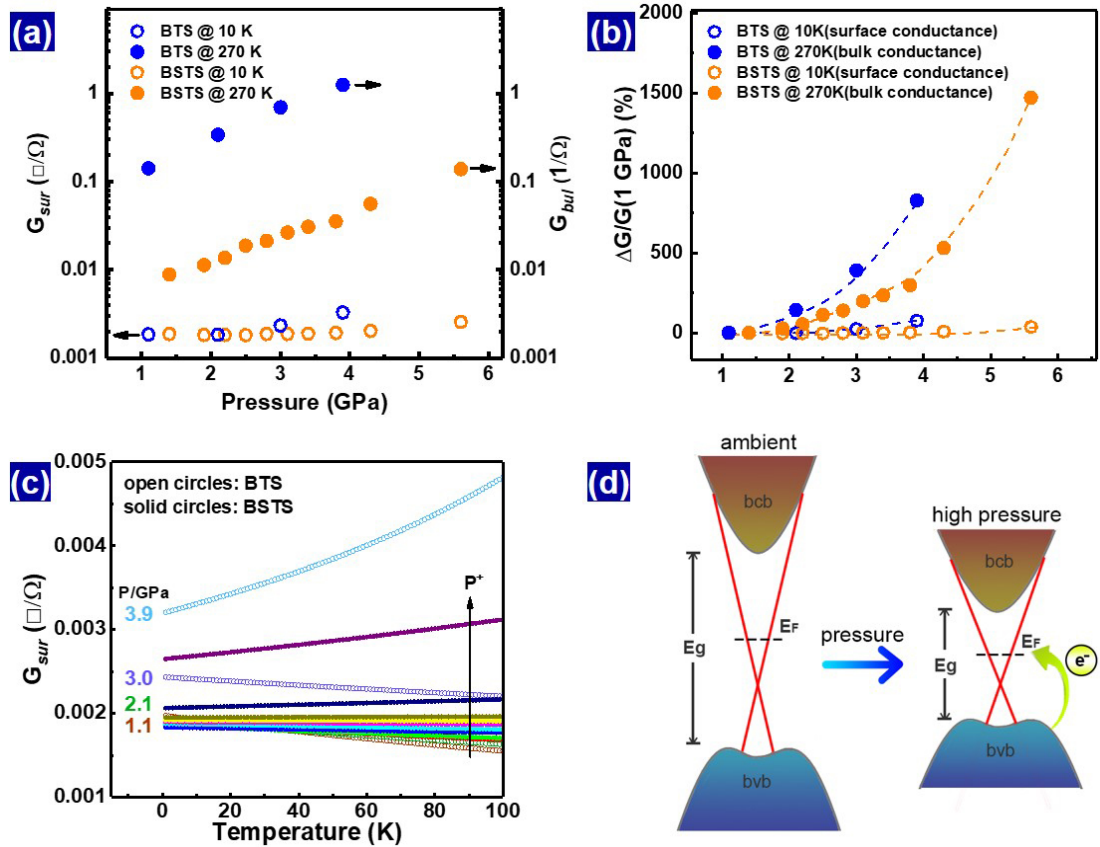

**Figure 3 Summary of the temperature and pressure dependent conductances for both BTS and BSTS.** (a) shows that the high temperature conductance converted from the measured resistance data, dominated by the bulk states, increases by more than an order of magnitude under pressure for both materials, while at low temperatures, (*i.e.* 10 K), where the surface state conductance dominates, the conductances are relatively independent of pressure. (b) illustrates the relative

changes in pressure-dependent conductances in specific, plotted as the ratio of the conductance change at pressure to the conductance at 1 GPa ($\Delta G = G(P) - G(1\ GPa)$), for both TIs, at both high and low temperatures. (c) shows the temperature dependence of the surface state conductances. The data for BSTS are essentially independent of pressure, while they do depend on pressure, somewhat, for BTS. This is due to the smaller band gap for the latter material and the dramatic decrease in the band gap with pressure for both materials, especially BTS, causing excitation of bulk carriers into and out of the surface states at intermediate temperatures. (d) A schematic of the electronic structures for optimally doped BTS and BSTS in the vicinity of the gamma point in the Brillouin Zone under pressure. (Note that this schematic is faithful to the electronic system for BSTS but that the Dirac crossing for BTS is within the notch in the valence band dispersion for that material, not relevant for the considerations here.)

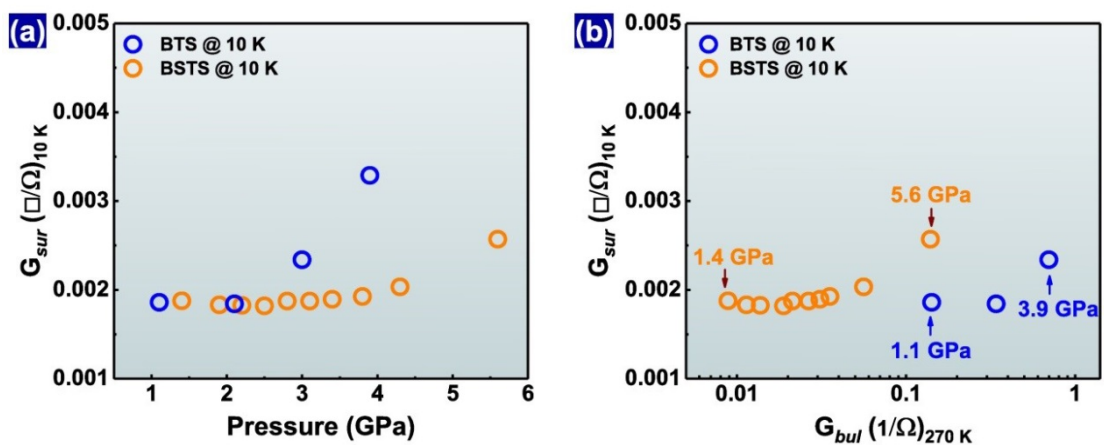

**Figure 4 The independence of surface state conductance and bulk conductance at low temperatures in topological insulators, which are converted from measured**

**resistance data. (a)** Pressure dependence of the surface state conductances in BTS and BSTS at 10 K. **(b)** The surface state conductance at 10 K plotted vs. the bulk state conductance at 270 K for both BTS and BSTS. Note the linear-log scales for figure (b). These representations show that in the low temperature and low pressure regime, where the bulk conductances are smallest, the surface state conductance is independent of both the bulk state conductance and the material studied.

**Supplementary Information for "Independence of surface state and bulk conductance in three-dimensional topological insulators"**

Shu Cai[1,4], Jing Guo[1], Vladimir A. Sidorov[3], Yazhou Zhou[1], Honghong Wang[1,4], Gongchang Lin[1,4], Xiaodong Li[5], Yanchuan Li[5], Ke Yang[6], Aiguo Li[6], Qi Wu[1], Jiangping Hu[1,4], S. K. Kushwaha[2], Robert J Cava[2]†, and Liling Sun[1,4]†

[1]Institute of Physics and Beijing National Laboratory for Condensed Matter Physics, Chinese Academy of Sciences, Beijing 100190, China

[2]Department of Chemistry, Princeton University, Princeton, New Jersey 08544, USA

[3]Institute for High Pressure Physics, Russian Academy of Sciences, 142190 Troitsk, Moscow, Russia

[4]University of Chinese Academy of Sciences, Beijing 100190, China

[5]Institute of High Energy Physics, Chinese Academy of Sciences, Beijing 100049, China

[6]Shanghai Synchrotron Radiation Facilities, Shanghai Institute of Applied Physics, Chinese Academy of Sciences, Shanghai 201204, China

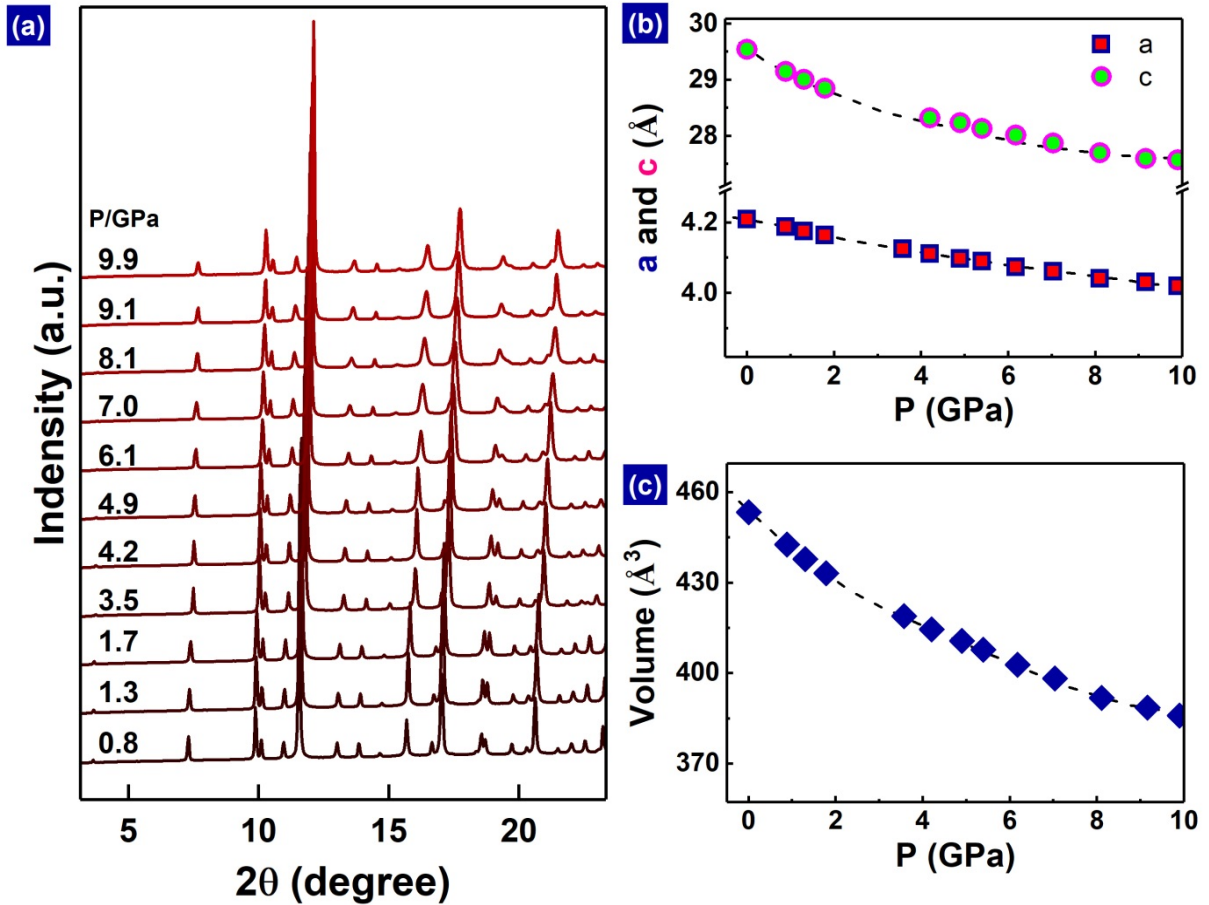

**Figure S1 High pressure X-ray diffraction study of BSTS, showing that it undergoes no structural phase transitions in the pressure range of these experiments**. (a) Synchrotron powder X-ray diffraction patterns of BSTS at different pressures between 0.8 and 9.9 GPa, showing the maintenance of the tetradymite-type diffraction pattern. (b) and (c) The pressure dependence of the lattice parameters and volume for BSTS.

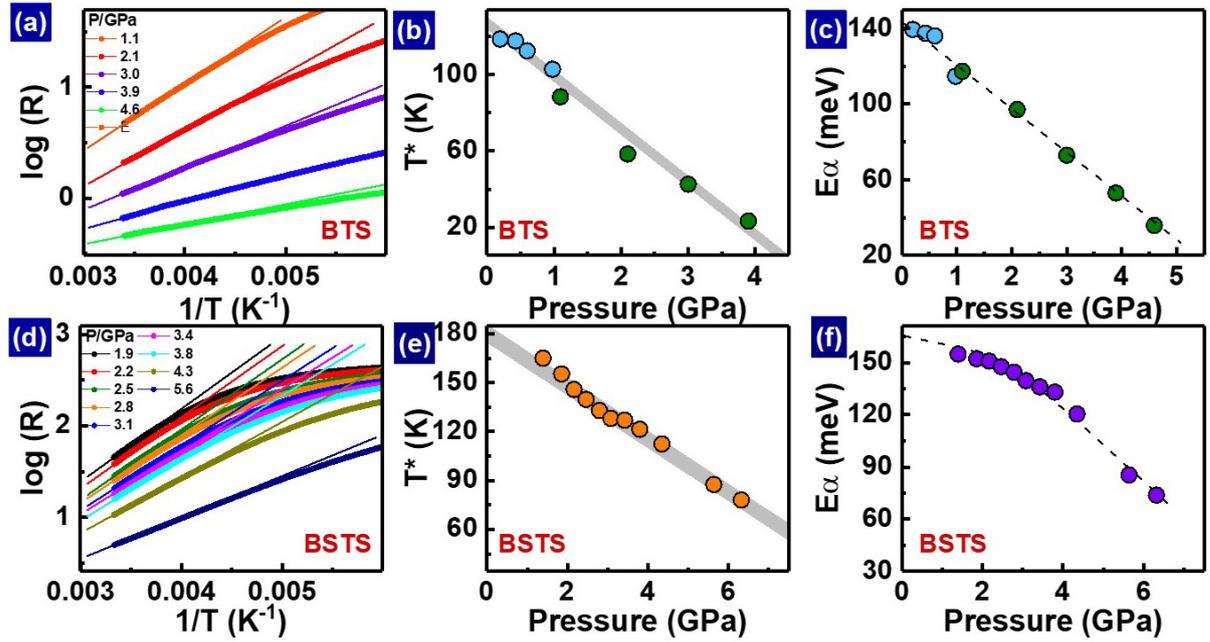

**Figure S2 The temperature-dependent resistances for crystals of the optimally doped topological insulators BTS and BSTS as a function of pressure.** (a) and (d) Arrhenius plots of the temperature dependence of the resistance at different pressures for the two materials. (b) and (e) Pressure dependence of the crossover temperatures $T^*$. (c) and (f) Pressure dependence of the resistance activation energies $E_\alpha$.

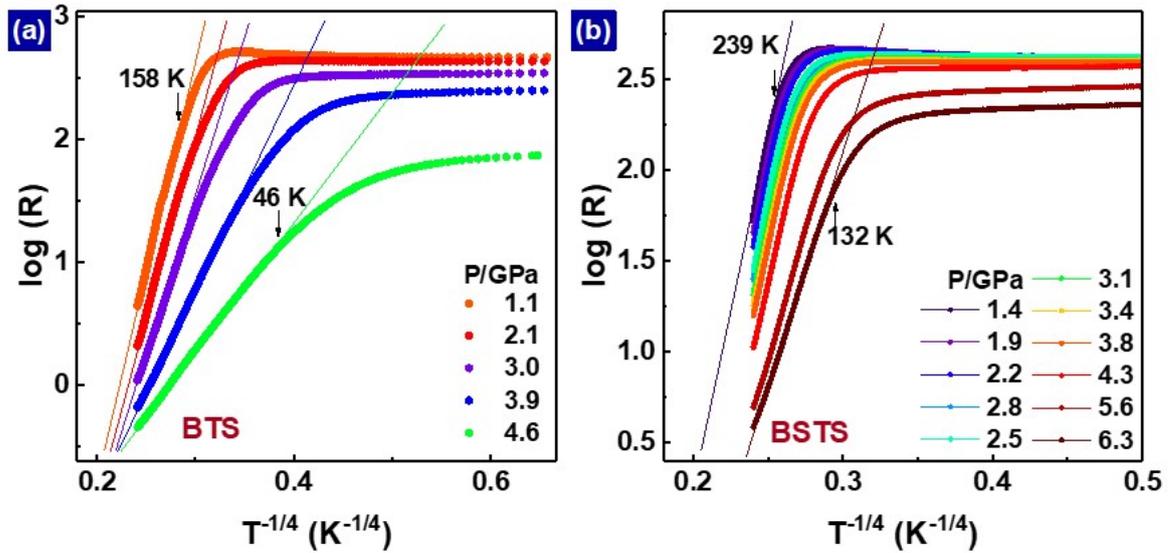

**Figure S3 Variable-Range-Hopping (VRH) plots of the temperature versus resistance at different pressures for the two materials.** The solids are experimental data and the lines are the linear fit to the data measured at high temperatures. It shows that the VRH fits can describe our experimental data well.

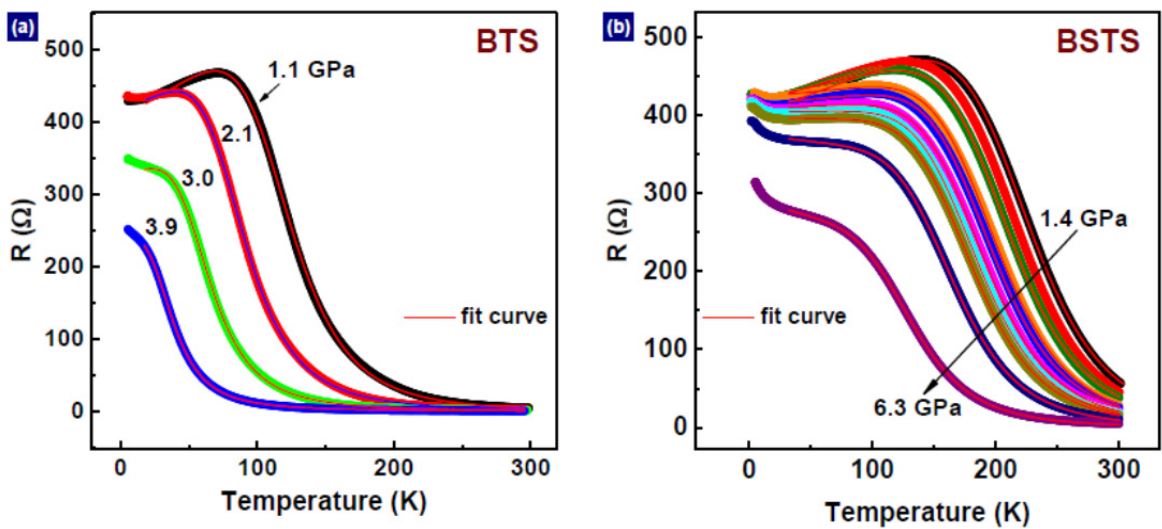

**Figure S4 Resistance as a function of temperature obtained at different pressures for BTS and BSTS samples.** The solids are experimental data and the red lines are the fit to the data with the two channel model.

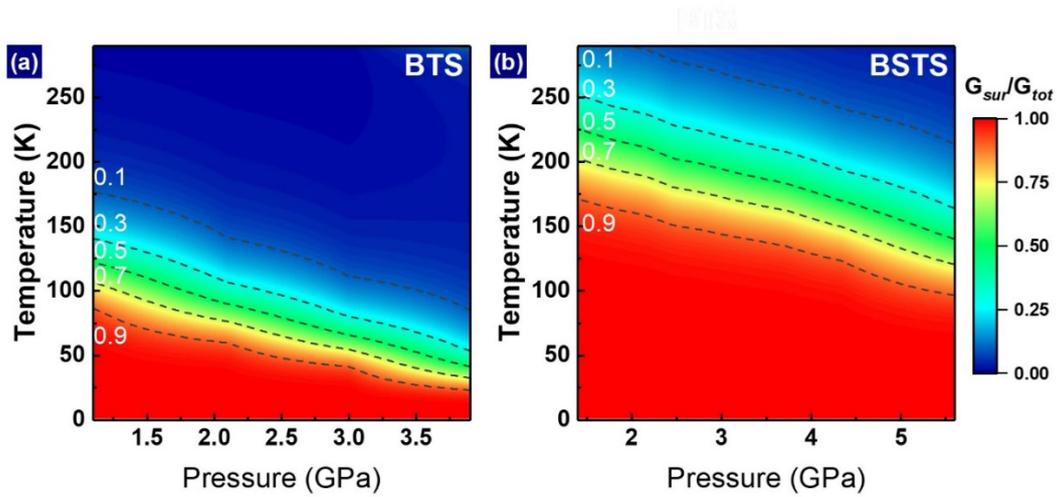

**Figure S5 Pressure dependence of the ratio of $G_{sur}/G_{tot}$ (shown on a color scale) for BTS and BSTS**. $G_{sur}$ and $G_{tot}$ are obtained from the bulk and surface conductances averaged from measurements for different runs, displaying that the pressure-induced decrease in the bulk activation energy influences the temperature of dominance of the TSS for both TIs. If we define the values of $G_{sur}/G_{tot}$ above 0.9 (red area) and below 0.1 (blue area) as the regimes where the metallic surface states and the insulating bulk states dominate respectively, then the similarities and differences between BTS and BSTS are straightforwardly seen through the color coding. Conservatively speaking, below 10 K at ambient pressures the surface states are sufficiently dominant in both materials to allow for study of their properties.

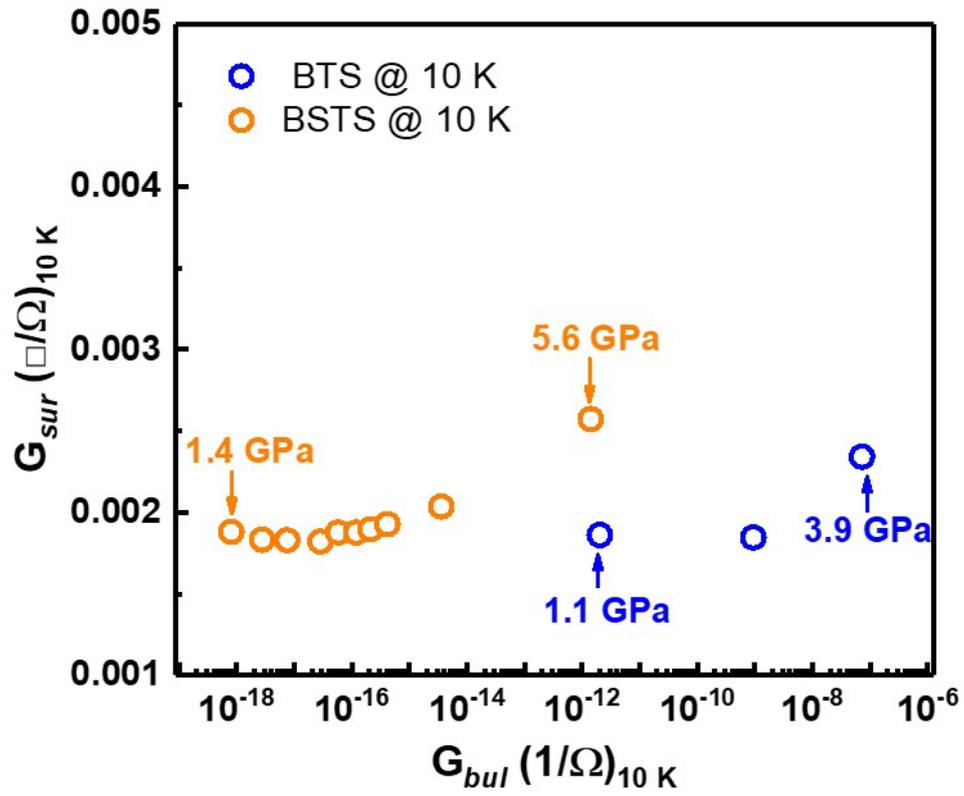

**Figure S6 Conductance of surface states as a function of the conductance of the bulk states extrapolated down to 10 K from the conductance starting at 270 K.** The orange circles represent the data obtained from the BSTS sample, while the blue circles stand for the data obtained from the BTS sample.